# TEMPERATURE DEPENDENCE OF THE PROBABILITY OF "SMALL HEATING" AND TOTAL LOSSES OF UCNs ON THE SURFACE OF FOMBLIN OILS OF DIFFERENT MOLECULAR MASS

S.M. Cherniavsky[1], E.V. Lychagin[2], A.Yu. Muzychka[2], G.V. Nekhaev[2], V.V. Nesvizhevsky[3], S. Reynaud[4], A.V. Strelkov[2], K. Turlybekuly[2]

[1]National Research Center Kurchatov Institute, 123182, Moscow, Russia
[2]Joint Institute for Nuclear Research, 141980, Dubna, Russia
[3]Institut Laue-Langevin, 38042, Grenoble, France
[4]Laboratoire Kastler Brossel, UPMN-Sorbonne Universities, CNRS, ENS-PSL Research University, College de France, Campus Jussieu, 75252, Paris, France

**Abstract**

We measured the temperature dependence of the probability of small heating and total losses of UCNs on the PFPE Fomblin Y surface with various molecular mass $M_{\bar{W}}$ (2800, 3300, 6500 amu) in the temperature range of 100-300 K. The probability of small heating sharply decreases with increasing $M_{\bar{W}}$ and decreasing temperature. The probability of total loss weakly decreases with decreasing temperature and takes the minimum value at $M_{\bar{W}} = 3300$ amu. As this oil provides a homogeneous surface with minimal probabilities of small heating and total losses of UCNs, it is the preferred candidate for experiments on measuring the neutron lifetime.

**Introduction**

In experiments with ultracold neutrons (UCNs), hydrogen-free oil Fomblin Y-HVAC 18/8 (perfluoropolyether oil (PFPE) Fomblin Y) with a mean molecular mass $M_{\bar{W}}$ =2800 amu is often used to cover the walls of neutron traps [1-3]. Being liquid at room temperature, it quickly loses its fluidity when cooled and passes into the glass phase at a temperature of ~196 K. The surface covered with a Fomblin layer is uniform and has a relatively high critical energy for UCNs: $E_{lim}$ =106.5 neV. The coefficient of total UCN loss via the capture channels and inelastic scattering on the walls of such a trap at room temperature is $\eta_{tot} \approx 2.4 \cdot 10^{-5}$ [4], which is significantly lower than the values typical for most other substances and materials. Extremely low pressure of saturated vapors of this oil guarantees the absence of noticeable mass transfer, which means the absence of the loss of UCNs on residual vapors and the stability of the transmission of the separating foils in experimental installations. Thus, a simple deposition of Fomblin on the surface of a closed vessel allows one to obtain an UCN trap with a storage time very close to the neutron lifetime due to the $\beta$-decay $\tau_\beta$ [5]. With the use of this oil, a number of experiments [6-11] were performed on the precision measurement of $\tau_\beta$ by the method of UCN storage.

However, the lack of a systematic study of the mechanism of interaction of UCNs with the surface of Fomblin oil creates significant problems for the analysis of experimental data. The mechanism of inelastic scattering of UCNs on the Fomblin surface, which inevitably affects the accuracy of the above experiments, has not yet been fully studied. The C, O, F nuclei entering the structure of the oil molecule have very small neutron capture cross sections. Therefore, it was originally assumed that the main contribution to the $\eta_{tot}$ value comes from inelastic scattering of UCNs into the range of thermal energies; the scattering occurs on collective vibrations of atoms located in the near-surface oil layer of ~30 nm deep, available for UCNs [7]. However, the calculated value of the total UCN loss coefficient obtained from experiments on the transmission of neutrons with a velocity of 9 m/s through the bulk layer of oil turned out to be 3 times lower than the value obtained from UCN storage experiments [12-13]. The difference can probably be explained by additional oscillation modes of the atoms, which appear exclusively on the surface of Fomblin oil.

In 1999, it was shown that in collisions with the surface of various substances, UCNs experience a "small heating" – specific inelastic scattering, leading to an increase in UCN energy by about two times [14-20]. This effect was also found on Fomblin Y-HVAC 18/8 oil [14-15, 19-21]. The most accurate measurements of this effect [22] showed that the probability of small heating of UCNs with an initial energy of 30 neV to the energy range of 35-140 neV turns out to be $\sim 10^{-5}$ per collision with an oil surface at room temperature. As the oil cools, the probability of small heating drops and becomes vanishingly small at the pour loss point temperature of Fomblin oil $T_{pour} = 231\ K$. This effect leads to a change not only in the energy spectrum of UCNs accumulated in the trap, but also in the spectrum of inelastically scattered neutrons leaving the trap, which makes it necessary to introduce additional systematic corrections in the already performed experiments on measuring the neutron lifetime $\tau_\beta$.

As a possible explanation of the effect of small heating of UCNs on the Fomblin surface, UCN scattering on capillary surface waves was considered [23-25]. There are also other hypotheses explaining the small heating of UCNs on the surface of hydrogen-free oil, for example, the scattering of UCNs on near-surface mesoscopic objects such as nanodroplets [26-27]. The experimental data on the spectrum of UCNs that have experienced "small heating" are not sufficient to give preference to one model or another, and to make corrections to the $\tau_\beta$ experiments. Attempts to eliminate the effect of small heating and reduce the probability of total losses of UCNs by cooling the trap to temperatures below 240 K have led to cracking and peeling of the oil coating [4, 12]. It should be noted that the best time of storage of UCNs was obtained in an experiment to measure $\tau_\beta$ in cryotraps covered with so-called low-temperature PFPE [28]. Unlike Fomblin oil, it hardens at a significantly lower temperature (80 K). At the same time, a detailed study of this oil with the aim of evaluating the degree of uniformity of the coating and the losses of UCNs on its surface as well as the reproducibility of the coating properties has not been carried out, which significantly reduced the value of the results obtained.

The measurements presented in the present paper provide the necessary data to make corrections to the performed experiments on measuring the neutron lifetime. We report the probabilities of small heating and total losses of UCNs on PFPE oil surfaces of one Fomblin Y type with different molecular masses $M_{\overline{W}}$ (2800, 3300, 6500 amu) in the temperature range of 100-300 K. This should allow one to find a new oil for neutron traps, for carrying out experiments free from systematic corrections for small heating of UCNs and also free from the problem of the Fomblin oil coating instability.

**Experimental setup**

A big gravitational spectrometer (BGS) was built specially to study the effect of small heating of UCNs on the surface of any substances and materials. Previously, we used it to measure the effect of small heating of UCNs on Fomblin Y with $M_{\overline{W}} = 2800$ amu. A detailed description of the principle of operation and design of BGS is given in [22].

The BGS allows forming a narrow spectral line of UCNs with an energy of 30 neV and observing small heating of UCNs to the energy range of 35-140 neV. The installation scheme is shown in Fig. 1. The spectrometer is a closed vessel for storage of UCNs made of copper with a diameter of 60 cm and height of 200 cm, in the upper part of which there is an absorber of polyethylene with a corrugated surface (5). At the bottom of the BGS, a potential barrier (2) is installed – a cylindrical surface with a diameter of 40 cm and a height of 35 cm separating the spectrometer space into two storage volumes – internal and external. The internal volume is designed for the accumulation and storage of UCNs, the energy of which does not allow them to rise above a height of 35 cm relative to the bottom of the BGS and pass into the external volume. The UCNs enter the internal volume via the input neutron guide (10), the upper edge of which is located 25 cm above the bottom of the storage volume, which determines the lower boundary of the initial UCN spectrum. The entrance shutter (3) locks the input neutron guide. The flux of UCNs at the bottom of the internal volume of the BGS is monitored using a monitor detector (4). The exit of UCNs from the storage volume into the monitor detector is closed by a flap in the form of a rising cover, in which there is a calibrated orifice with a diameter of 5 mm (8). The device «elevator» (9) allows placing samples at the bottom of the internal volume of the spectrometer without breaking the vacuum. The upper boundary of the UCN spectrum stored in the internal volume of the BGS is formed by a polyethylene absorber (5) installed at a height of 32 cm relative to the bottom of the spectrometer, which makes it impossible to pass UCNs from the internal to the external volume. If, because of "small heating", UCNs acquire energy that allows them to overcome the potential barrier (2), they will flow into the external volume of the spectrometer and be counted in the detector (6). The shutter (7) allows blocking the leakage of UCNs to the detector (6). By the ratio of the detector (6) count and the monitor (4) count, one can determine the probability of small heating of UCNs. Measurements of the UCN storage times in the internal volume of the BGS with a sample and without a sample make it possible to determine the total probability of UCN loss on the sample.

In order to improve the storage time of UCNs, all the fixed surfaces of the BGS internal volume were coated with a brush by a thin layer of Fomblin Y-HVAC 140/13 [1] with the molecular mass $M_{\overline{W}} = 6500$ amu. The thickness of the layer, estimated by the oil consumption, is 1-10 μm. The moving parts of the shutter were coated with lubricant based on the same oil in order to avoid possible splashing of the oil and contamination of the surface of the separating foils (separating the vacuum of the transport neutron guide and storage volume) with oil.

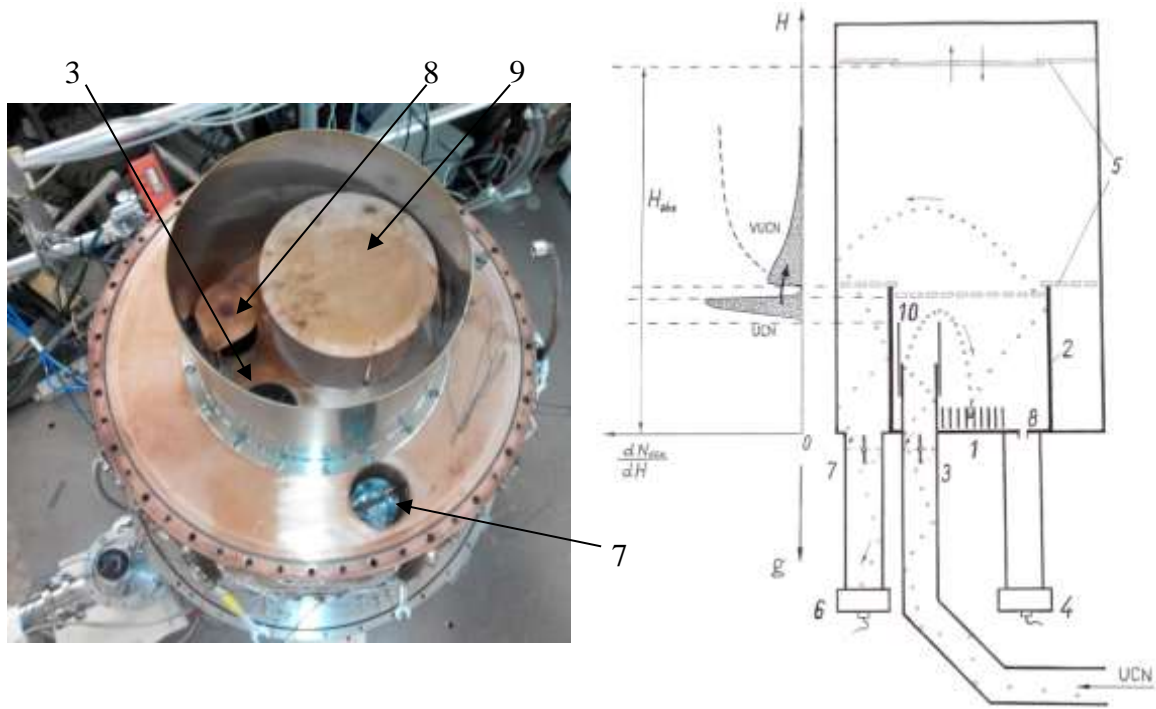

*Fig. 1. Installation scheme. 1 – sample, 2 – gravitational barrier, 3 – entrance shutter, 4 – monitor, 5 – absorber, 6 – detector, 7 – exit shutter, 8 – calibration orifice, 9 — "elevator" for samples, 10 – entrance neutron guide.*

**Samples**

Surfaces coated with Fomblin Y-HVAC oils with three different molecular masses $M_{\overline{W}}$: 18/8 – 2800 amu (F-2800), 25/9 – 3300 amu (F-3300) and 140/13 – 6500 amu (F-6500) [1, 29] were used as samples (notations 18/8, 25/9, 140/13 are introduced by Solvay, the producer of Fomblin). These oil samples have approximately the same physical properties at room temperature, with the exception of saturated vapor pressure and viscosity, which vary nonlinearly with the mean molecular mass of the oil (Table 1).

*Table 1. Physical parameters of Fomblin Y-HVAC at room temperature [30].*

| Oil type | $M_{\overline{W}}$, a.e. | Density, $g/cm^3$ | Heat capacity, $cal/g \cdot K$ | Coefficient of surface tension, $dyne/cm$ | Viscosity, $cm^2/s$ | $T_{pour}, K$ |
|---|---|---|---|---|---|---|
| 18/8 | 2800 | 1.89 | 0.24 | 20 | 1.9 | 231 |
| 25/9 | 3300 | 1.90 | 0.24 | 20 | 2.5 | 238 |
| 140/13 | 6500 | 1.92 | 0.24 | 20 | 15.08 | 250 |

Each type of oil was applied with a brush on the surface of aluminum foils with a thickness of 100 μm. Identical samples with a total area of 0,74 m² were assembled from the foils. The sample diameter is ~17 cm, height is 5 cm. Fig. 2 shows a photo of the sample.

For the similarity of the initial experimental conditions, we used oils directly as they were obtained from the manufacture, without any additional sample preparation procedures (without heating, without prior degassing, etc).

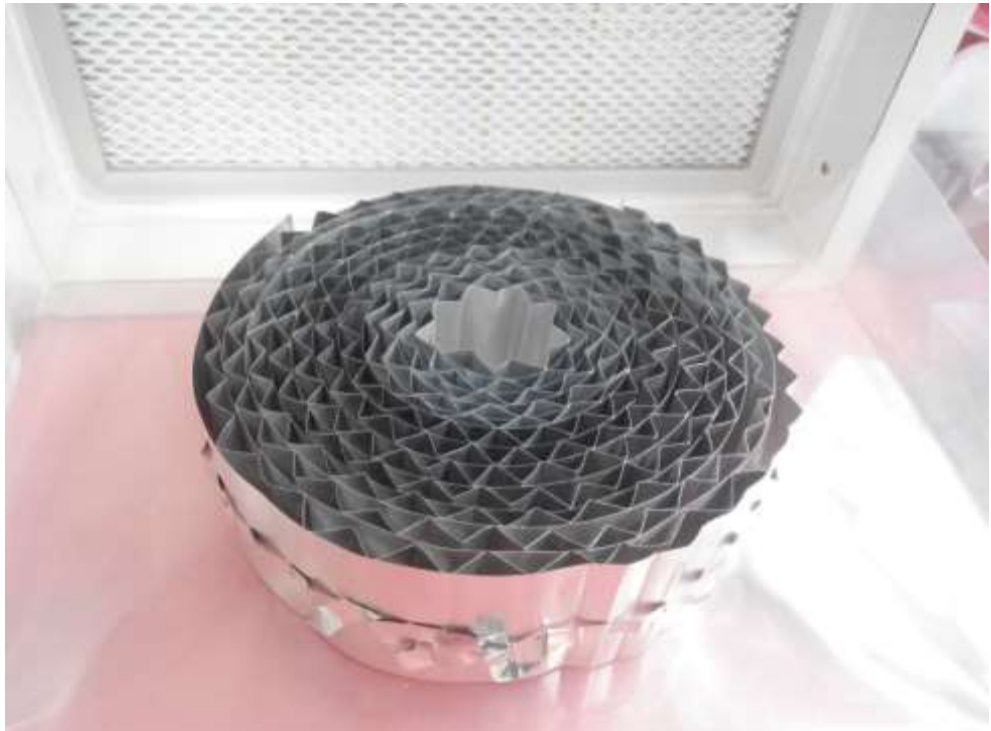

*Fig. 2. A photo of the sample.*

**Measurement technique**

For each sample, we measured the probability of small heating and the coefficient of total loss of UCNs in the temperature range of 100-300 K. The sample temperature was measured by a K-type thermocouple attached to the sample.

The cycle of measurement of the probability of small heating consists of the following steps: filling (150 s), cleaning (150 s), measuring the effect (600 s), emptying the trap (150 s). At all these stages, the exit shutter in front of the detector (7) is open, the shutter in front of the monitor detector is closed.

The top of the entrance neutron guide was always set at the height of 25 cm above the bottom of the BGS; during the spectrometer filling and cleaning, the polyethylene absorber is lowered to a height of 32 cm relative to the bottom of the BGS. Thus, within 150 s, the initial UCN spectrum of 25-32 neV is shaped in the internal volume of the spectrometer (within the measurement accuracy, we assume neutron rise in the Earth's gravitational field of 1 cm corresponds to its potential energy 1 neV).

After the completion of filling, the entrance shutter (3) closes and cleaning phase begins. Neutrons reaching the absorber quickly are lost in it or flow out of the external volume to the detector (6). As a result, the detector count rate decreases sharply to the background value (the detector background is $(9.6 \pm 0.7) \cdot 10^{-4}$ s$^{-1}$, the monitor background is $(3.70 \pm 0.13) \cdot 10^{-4}$ s$^{-1}$). The duration of cleaning $t_{clean} = 150\ s$ is chosen so that the number of UCNs remaining in the internal volume capable of reaching the absorber does not affect the measurement results. At the end of the cleaning procedure, the absorber rises to a height of $140\ cm$ relative to the bottom of the BGS spectrometer and the effect measurement stage begins. The rise of the absorber has no effect on the UCN spectrum formed in the internal volume and, accordingly, on the monitor, while the detector count increases rapidly and after some time becomes proportional to the flux of UCNs stored in the internal volume. Permanent generation of so-called VUCNs (vaporizing UCNs), neutrons with energy allowing them to leave the internal volume, explains this dependence. If in the course of inelastic scattering on the surface of the spectrometer or a sample, an UCN has acquired energy that allows it to reach the absorber, then the probability of its passage to the detector is suppressed by the probability of its loss in the absorber. Thus, the detector counts predominantly neutrons with energy in the range of 35-140 neV relative to the bottom of the spectrometer.

At the 900$^{th}$ second, the absorber is lowered to a height of 32 cm, and the detector count rate drops again to the background value.

As other than sample surfaces of the spectrometer are also coated with F-6500 oil, to calculate the probability $P_{VUCN}$ of small heating on the F-6500 sample, we used the following expression [22]:

$$P_{VUCN} = \frac{N_{det}}{N_{mon} \varepsilon} \frac{S_{mon}}{S_{sample}} , \qquad (1)$$

where $N_{det}$ and $N_{mon}$ are, respectively, the counts of the detector and monitor at the stage of measuring the effect, $\varepsilon$ is the average VUCN detection efficiency, $S_{mon}$ and $S_{sample}$ are the effective areas of the monitor orifice and the surface of the spectrometer with the sample.

For other samples, this expression was modified to exclude small heating on the surface of the walls of the storage volume.

The detection efficiency is determined by the ratio of the VUCN emptying time and the lifetime of VUCNs in the spectrometer. The latter is mainly determined by the losses in the external storage volume and is weakly dependent on temperature. The procedure for measuring and calculating the efficiency is described in detail in [22].

To measure the total UCN loss coefficient, the UCN storage times in the internal volume of BGS with and without a sample were measured. To determine the storage time, the following measurement diagram was implemented: filling (150 s, the absorber is at a height of 32 cm), cleaning (150 s, the entrance shutter is closed), storage (the absorber is at a height of 37 cm, to exclude the possible influence of «uncleaned» neutrons on the measurement results), emptying (80 s, the monitor shutter is open). At the end of the cycle, the monitor shutter closes and the absorber returns to its original position. Statistics are collected by repeating pairs of cycles with storage times $\Delta t_{st1}$ and $\Delta t_{st2}$, thus, the monitors counts $N(\Delta t_{st1})$ and $N(\Delta t_{st2})$ are determined. The storage time is calculated using the formula:

$$\tau_{st} = \frac{\Delta t_{st2} - \Delta t_{st2}}{\ln(N(\Delta t_{st2})) - \ln(N(\Delta t_{st1}))} \ . \tag{2}$$

The storage time of UCNs in the spectrometer with the sample $\tau_{st}(sample, T)$ can be calculated as follows:

$$\frac{1}{\tau_{st}(sample,T)} = \eta_{sample}(T)\gamma_{sample} + \eta_{sp}(T)\gamma_{sp} + \frac{1}{\tau_\beta} + \frac{1}{\tau_{slits}}, \tag{3}$$

where $\eta_{sample}(T)$ is the total loss coefficient for the sample at temperature $T$, $\tau_\beta$ is the neutron lifetime ($\tau_\beta$=880.2±1.0 [5]), $\tau_{slits}$ is the characteristic time of neutron loss in the slits of the shutters, the sample elevator and the monitor orifice, $\gamma_{sample}$ and $\gamma_{sp}$ are "gamma-functions", or the average effective frequency of UCN collisions with the sample and with the empty spectrometer with the elevator raised, respectively [22].

Taking into account the narrowness of the shaped UCN spectrum, to determine the loss coefficients, the values of "gamma-functions" were calculated for the average neutron energy corresponding to the height of 30 cm: $\gamma_{sample} = 13.0 \ s^{-1}$, $\gamma_{sp} = 7.4 \ s^{-1}$

Thus, by conducting two consecutive measurements of the UCN storage time in the spectrometer with and without a sample, using formula (3), one can obtain the value of the coefficient of total losses for the sample.

**Results of measurements and discussion of the results**

Measurements were carried out at slowly changing temperature. The accuracy of temperature settings in all measurements except measurements at room temperature is ±2,5 K.

The results of measurements of the temperature dependence of the probability of small heating of UCNs are presented in Fig. 3. $P_{VUCN}$ for all samples decreased rapidly as the temperature decreases and reaches <1·10⁻⁷ per collision at the pour point $T_{pour}$. We note that the measured $P_{VUCN}$ values for F-2800 and F-3300 samples coincide within the statistical accuracy, despite the difference in average molecular weights and viscosities. At the same time, $P_{VUCN}$ for sample F-6500, already at room temperature, is two times lower than for the other two samples while the molecular mass is two times different and the viscosity is almost tenfold different. This difference in $P_{VUCN}$ grows as the temperature decreases.

From the point of view of experiments on measuring the neutron lifetime, the corrections for the effect of «small heating» to $\tau_\beta$ become smaller than 0.2 s at a temperature of 270 K for samples F-2800 and F-3300 and 285 K for F-6500, while for sample F-2800 a correction at room temperature can reach 5 s [25].

The results of measuring the UCN loss coefficient are represented in Fig. 6. The observed drop in the total loss coefficient when the samples are cooled to a temperature of 270 K is caused to a large extend by a drop in the contribution of the small heating to the value $\eta_{tot}$. Indeed, in this experiment, the small heating of UCNs stored in the internal volume of the BGS spectrometer leads to their transition to the external volume, which is equivalent to their loss. Therefore, in this case, the expression for the total probability of neutron losses with an energy $E_0$ is:

$$\mu_{tot}(E_0) = \eta_0(T) f\left(\sqrt{\frac{E_0}{E_{lim}}}\right) + P_{VUCN}(T, E_0), \tag{4}$$

where $f(x) = \frac{2}{x^2}(\arcsin(x) - x\sqrt{1-x^2})$, $\eta_0(T)$ is the total UCN loss coefficient due to the radiative neutron capture and their inelastic scattering to the thermal region, $E_{lim}$ is the boundary energy of the potential barrier (2). The obtained loss coefficient $\eta(T)$ is related to $\eta_0(T)$ as follows:

$$\eta(T) = \eta_0(T) + P_{VUCN}(T, E_0)/f\left(\sqrt{\frac{E_0}{E_{lim}}}\right). \tag{5}$$

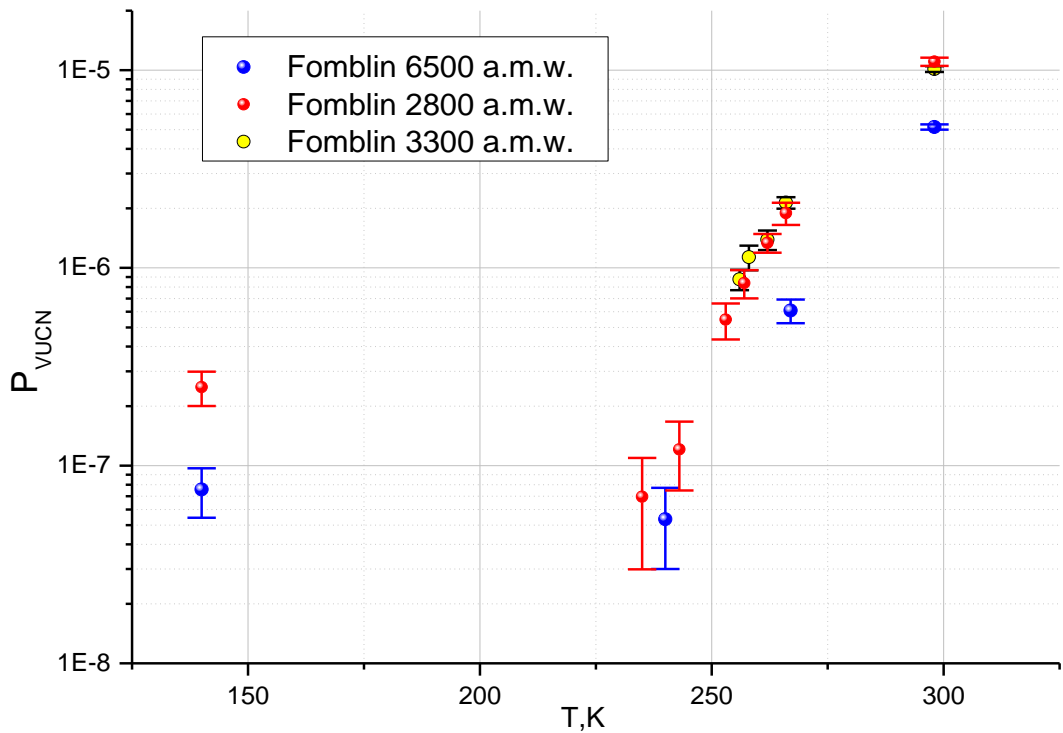

*Fig. 3. Temperature dependence of the probability of small heating to the energy range of 35÷140 neV for Fomblin Y oils.*

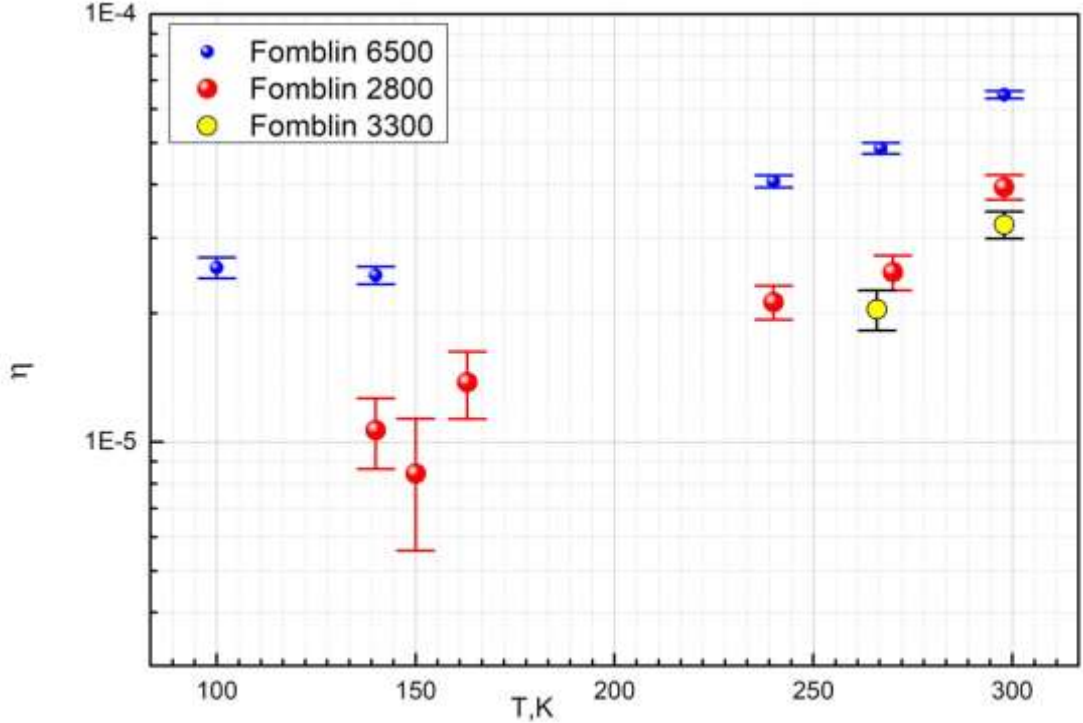

*Fig. 4. Temperature dependence of the loss coefficient for Fomblin Y oils.*

At temperatures below 270 K, the value of the total loss coefficient $\eta(T)$ practically coincides with $\eta_0(T)$. However, at temperatures around 170 K, the value of $\eta(T)$ ceases to decrease. In this regard, a visual observation was made of the behavior of thin oil coatings during cooling in the temperature range of 130-300 K. The coatings were applied to the polished surface of a copper cylinder, which plays the role of a cooling pipe from tubes through which liquid nitrogen vapor was circulated. The cooling pipe was placed inside an evacuated chamber with a glass window that allows monitoring the state of the oil coating. The chamber was pumped out with an oil-free pump to a pressure of $3 \cdot 10^{-3}$ bar. The cooling rate in the temperature range of 200-300 K was 4-6 K/min, and at temperatures below 200 K it is 1-2 K/min.

For all types of oil studied at the specified cooling rate, with a layer thickness of 10-200 μm, starting at temperatures of 165-175 K, spontaneous cracking of coatings was observed. With further cooling, the entire observation area was covered with small cracks. Fig. 5 (a) shows a typical picture of the destruction of the F-2800 oil layer. The layers of UT-18 Fomblin grease begin to crack at a temperature of 130 K, completely covered with small cracks at 120 K as shown in Fig. 5(б).

Thus, despite the fact that for each of the studied samples, the value of the total UCN loss coefficient upon reflection from oil coatings reaches its minimum at temperatures of 100-170 K, the inhomogeneous structure that arises is a significant problem for experiments on the precise measurement of the neutron lifetime, which assume the similarity of trap wall surfaces.

At room temperature and at a temperature of $T$=270 K, sample F-3300 showed the best storage time, which makes this oil potentially interesting for further use in experiments to measure the neutron lifetime: at a temperature of $T$ =270 K, the surface of the oil remains homogeneous, and the effect of small heating can be neglected at the level of accuracy of 0.2 s for $\tau_\beta$.

The sample of F-6500 oil at room temperature has a very high viscosity, which probably makes it difficult for the gases dissolved in it to exit. Perhaps preheating in vacuum at a temperature of 360-380 K will allow the oil layer to be degassing and to noticeably improve values of the total UCN loss probability.

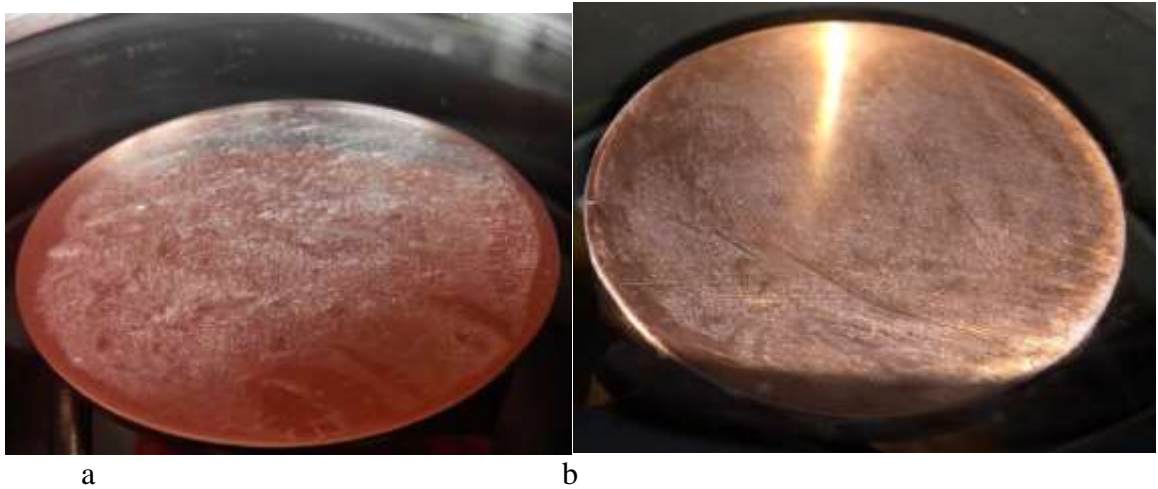

a                                                b

*Fig. 5. a) A layer of F-2800 oil at a temperature of 140 K, b) a layer of lubricant U-18 at a temperature of 120 K.*

**Conclusion**

Using the dedicated BGS spectrometer, we investigated the probability of small heating and total losses of UCNs at their collisions with PFPE Fomblin Y surfaces with various molecular masses $M_{\bar{W}}$ (2800, 3300, 6500 amu) in a broad temperature range of 100-300 K. For all types of oil, the probability of small heating sharply decreases with increasing $M_{\bar{W}}$ and decreasing temperature. The probability of total loss weakly decreases with decreasing temperature and takes the minimum value at $M_{\bar{W}} = 3300$ amu. Since this oil provides a homogeneous surface with minimal probabilities of small heating and total losses of UCNs at a temperature of 270-250 K, it is the preferred candidate for experiments on measuring the neutron lifetime. We investigated reasons for poor reproducibility of experimental results at low temperatures. Visual observations showed that for all types of oil studied spontaneous cracking of coatings is observed at temperatures of 165-175 K. With further cooling, the entire observation area was covered with small cracks. Thus, despite the fact that for each of the studied samples, the value of the total UCN loss coefficient upon reflection from oil coatings reaches its minimum at temperatures of 100-170 K, the inhomogeneous structure that arises is a significant problem for experiments on the precise measurement of the neutron lifetime, which assume the similarity of trap wall surfaces.